\documentclass{article}
\usepackage{amssymb}

\usepackage{graphicx}
\usepackage{amsmath}

\newtheorem{theorem}{Theorem}

\input{tcilatex}

\begin{document}

\title{Betting on the Outcomes of Measurements: \\
A Bayesian Theory of Quantum Probability}
\author{Itamar Pitowsky \\
%EndAName
Department of Philosophy, the Hebrew University,\\
Mount Scopus, Jerusalem 91905, Israel. itamarp@vms.huji.ac.il}
\maketitle

\begin{abstract}
We develop a systematic approach to quantum probability as a theory of
rational betting in quantum gambles. In these games of chance the agent is
betting in advance on the outcomes of several (finitely many) incompatible
measurements. One of the measurements is subsequently chosen and performed
and the money placed on the other measurements is returned to the agent. We
show how the rules of rational betting imply all the interesting features of
quantum probability, even in such finite gambles. These include the
uncertainty principle and the violation of Bell's inequality among others.
Quantum gambles are closely related to quantum logic and provide a new
semantics to it. We conclude with a philosophical discussion on the
interpretation of quantum mechanics.
\end{abstract}

\section{Quantum Gambles}

\subsection{The Gamble}

The Bayesian approach takes probability to be a measure of ignorance,
reflecting our state of knowledge and not merely the state of the world. It
follows Ramsey's contention that ``we have the authority both of ordinary
language and of many great thinkers for discussing under the heading of
probability ... the logic of partial belief'' (Ramsey 1926, p. 55). Here we
shall assume, furthermore, that probabilistic beliefs are expressed in
rational betting behavior: ``The old-established way of measuring a person's
belief ... by proposing a bet, and see what are the lowest odds which he
will accept, is fundamentally sound''\footnote{%
Ramsey, 1926, p. 68. This simple scheme suffers from various weaknesses, and
better ways to associate epistemic probabilities with gambling have been
developed (de Finetti, 1972). Any one of de Finetti's schemes can serve our
purpose. For a more sophisticated way to associate probability and utility
see Savage (1954 )}. My aim is to provide an account of the peculiarities of
quantum probability in this framework. The approach is intimately related to
the foundational work on quantum information by Barnum et al (2000), Fuchs
(2001), Schack, Brun and Caves (2001) and Caves, Fuchs and Schack (2002)..

For the purpose of analyzing quantum probability we shall consider \textit{%
quantum gambles.} Each quantum gamble has four stages:

1. A \textit{single} physical system is prepared by a method known to
everybody.

2. A \textit{finite} set $\mathcal{M}$ of incompatible measurements is
announced by the bookie, and the agent is asked to place bets on possible
outcomes of each one of them.

3. One of the measurements in the set $\mathcal{M}$ is chosen by the bookie
and the money placed on all other measurements is promptly returned to the
agent.

4. The chosen measurement is performed and the agent gains or looses in
accordance with his bet on that measurement.

We do not assume that the agent who participates in the game knows quantum
theory. We do assume that after the second stage, when the set of
measurements is announced, the agent is aware of the possible outcomes of
each one of the measurements, and also of the relations (if any) between the
outcomes of different measurements in the set $\mathcal{M}$. Let me make
these assumptions precise. For the sake of simplicity we shall only consider
measurements with a finite set of possible outcomes. Let $A$ be an
observable with $n$ possible distinct outcomes $a_{1,}a_{2,...,}a_{n}$. With
each outcome corresponds an \textit{event} $E_{i}=\{A=a_{i}\}$, $i=1,2,...,n$%
, and these events generate a Boolean algebra which we shall denote by $%
\mathcal{B}=\left\langle E_{1},E_{2},...,E_{n}\right\rangle $. \textit{%
Subsequently we shall identify the observable }$A$\textit{\ with this
Boolean algebra.} Note that this is an unusual identification. It means that
we equate the observables $A$ and $f(A)$, whenever $f$ is a one-one function
defined on the eigenvalues of $A$. This step is justified since we are
interested in \textit{outcomes} and not their labels, hence the scale free
concept of observable. With this $\mathcal{M}$ is a finite family of finite
Boolean algebras. Our first assumption is that the agent knows the number of
possible distinct outcomes of each measurement in the set $\mathcal{M}$.

Our next assumption concerns the case where two measurements in the set $%
\mathcal{M}$ share some possible elements. For example, let $A,B,C$ be three
observables such that $[A,B]=0$, $[B,C]=0$, but $[A,C]\neq 0$. Consider the
two incompatible measurements, the first of $A$ and $B$ \ together and the
second of $B$ and $C$ together. If $\mathcal{B}_{1}$ is the Boolean algebra
generated by the outcomes of the first measurement and $\mathcal{B}_{2}$ of
the second, then $\mathcal{M}=\{\mathcal{B}_{1},\mathcal{B}_{2}\}$ and the
events $\{B=b_{i}\}$ are elements of both algebras, that is of $\mathcal{B}%
_{1}\cap \mathcal{B}_{2}$. We assume that the agent is aware of these facts
when he is placing his bets.

The smallest nontrivial case of this kind is depicted in figure 1. The graph
represents two Boolean algebras $\mathcal{B}_{1}=\left\langle
E_{1},E_{2},E_{3}\right\rangle $, $\mathcal{B}_{2}=\left\langle
E_{1},E_{4},E_{5}\right\rangle $ corresponding to the outcomes of two
incompatible measurements and they share a common event $E_{1}$. The
complement of $E_{1}$ denoted by $\overline{E}_{1}$ is identified as $%
E_{2}\cup E_{3}=E_{4}\cup E_{5}$. The edges in the graph represent the
partial order relations in each algebra from bottom to top. A realization of
these relations can be obtained by the system considered in Kochen and
Specker (1967): Let $S_{x}^{2}$, $S_{x`}^{2}$, $S_{y}^{2}$, $S_{y`}^{2}$, $%
S_{z}^{2}$ be the squared components of spin in the $x$, $x`$, $y$, $y`$, $z$
directions of a spin-1 (massive) particle, where $x,y,z$ and $x`,$ $y`,$ $z$
\ form two orthogonal triples of directions with the $z$-direction in
common. The operators $S_{x}^{2}$, $S_{y}^{2}$ and $S_{z}^{2}$ all commute,
and have eigenvalues $0,1$. They can be measured simultaneusly, and they
satisfy $S_{x}^{2}+S_{y}^{2}+S_{z}^{2}=2I$. Similar relations hold in the
other triple $x`,$ $y`,$ $z$. Hence, if we define $E_{1}=\{S_{z}^{2}=0\}$, $%
E_{2}=\{S_{x}^{2}=0\}$, $E_{3}=\{S_{y}^{2}=0\}$, $E_{4}=\{S_{x`}^{2}=0\}$, $%
E_{5}=\{S_{y`}^{2}=0\}$ we obtain the two Boolean algebras depicted in
figure 1.

\begin{center}
\FRAME{ftbpF}{2.879in}{2.1612in}{0pt}{}{}{bayes01.eps}{\special{language
"Scientific Word";type "GRAPHIC";maintain-aspect-ratio TRUE;display
"USEDEF";valid_file "F";width 2.879in;height 2.1612in;depth
0pt;original-width 8.534in;original-height 6.3866in;cropleft "0";croptop
"1";cropright "1";cropbottom "0";filename 'Bayes01.eps';file-properties
"XNPEU";}}
\end{center}

To sum, we assume that when the set of measurements $\mathcal{M}$ is
announced the agent is fully aware of the number of outcomes in each
measurement and of the relations between the Boolean algebras they generate.
In the spin-1 case just considered it means that the agent is aware of the
graph structure in figure 1. We shall refer in short to this background
knowledge as \textit{the logic of the gamble}. We assume no further
knowledge on the part of the agent, in particular, no knowledge of quantum
mechanics. Our purpose is to calculate the constraints on the probabilities
that a rational agent can place in such gambles.

\subsection{Methodological Interlude: Identity of Observables and
Operational Definitions}

Already at this stage one might object that the identity of observables in
quantum mechanics \textit{depends on probability}. Consider the case of the
operators $A,B,C$ such that $[A,B]=0$, $[B,C]=0$, but $[A,C]\neq 0$, and the
two incompatible measurements of $A$ together with $B$, and of $B$ together
with $C$. We are assuming that the agent is aware of the fact that the
events $\{B=b_{i}\}$ are the same in both measurements. However, the actual
procedure of measuring $B$ can be very different in the two cases, so how is
such awareness comes about? Indeed, the identity criterion for (our kind of)
observables is: \textit{Two procedures constitute measurements of the same
observable if for any given physical state (preparation) they yield
identical probability distribution over the set of possible outcomes}%
\footnote{%
In a deterministic world we would have a different criterion: \textit{Two
procedures constitute measurements of the same observable if for any given
physical state they yield identical outcomes. }We shall come back to this
criterion in section 3.2}\textit{. }It seems therefore that foreknowledge of
the probabilities is a necessary condition for defining the identities of
observables. But now we face a similar problem, how would one know when two 
\textit{states} are the same? Identical states can be prepared in ways that
are physically quite distinct. Well, \textit{two state preparations are the
same if for any given measurement they yield the same distribution of
outcomes}. A vicious circle.

There is nothing special about this circularity, a typical characteristic of
operational ``definitions'' (Putnam, 1965). In fact, one encounters a
similar problem in traditional probability theory in the interplay between
the identity of events and their probability. The way to proceed is to
remember that the point of the operational exercise is not to reduce the
theoretical objects of the theory to experiments, but to \textit{analyze }%
their meaning and their respective role in the theory. In this idealized and
nonreductive approach one takes the identity of one family of objects as
somehow given, and proceeds to recover the rest.

Consider how this is done in a recent article by Hardy (2001). Assuming that
the probabilities of quantum measurements are experimentally given as
relative frequencies, and assuming they satisfy certain relations, Hardy
derives the structure of the observables (that is, the Hilbert space). His
``solution'' to the problem of the identity of states, or preparations, is
simple. He stipulates that ``preparation'' corresponds to a position of a
certain dial, one dial position for each preparation. The problem is simply
avoided by idealizing it away.

Our approach is the mirror image of Hardy's. We are assuming that the
identities of the observables (and in particular, events) are given, and
proceed to recover the probabilities. This line of development is shared
with all traditional approaches to probability where the identity of the
events is invariably assumed to be given prior to the development of the
theory. It is, moreover, easy to think of an idealized story which would
cover our identity assumption. For example, in the three operator case $%
A,B,C $ mentioned above, we can imagine that the results of their
measurements are presented on three different dials. If $B$ is measured
together with $A$ then the $A$-dial and $B$-dial show the results; if $B$
and $C$ are measured together the $B$-dial and $C$-dial show the results.
Thus, fraud notwithstanding, the agent knows that he faces the measurement
of the same $B $ simply because the same gadget shows the outcome in both
cases.

\subsection{Rules of Gambling}

Our purpose is to calculate the constraints on the probabilities that a
rational agent can place in a quantum gamble $\mathcal{M}$. These
probabilities have the form $p(F\mid \mathcal{B})$ where $\mathcal{B}\in 
\mathcal{M}$ and$\ F\in $ $\mathcal{B}$. The elements $F\in $ $\cup _{%
\mathcal{B}\in \mathcal{M}}\mathcal{B}$ will be called simply ``events''.\ 
\textit{It is understood that an event is always given in the context of a
measurement }$\mathcal{B}\in \mathcal{M}$. The probability $p(F\mid \mathcal{%
B})$ is the degree of belief that the event $F$ occurs in the measurement $%
\mathcal{B}$. There are two rules of rational gambling, the first is
straightforward and the second more subtle.

\textbf{RULE\ 1:} \textit{For each measurement }$\mathcal{B}\in \mathcal{M}$ 
\textit{the function }$\ p(\cdot \mid \mathcal{B})$\textit{\ is a
probability distribution on }$\mathcal{B}$.

This follows directly from the classical Bayesian approach. Recall that
after the third stage in the quantum gamble the agent faces a bet on the
outcome of a single measurement. The situation at this stage is essentially
the same as a tossing of a coin or a casting of a dice. Hence, the
probability values assigned to the possible outcomes of the chosen
measurement should be coherent. In other words, they have to satisfy the
axioms of the probability calculus. The argument for that is that an agent
who fails to be coherent will be compelled by the bookie to place bets that
will cause him a sure loss (this is the ``Dutch Book'' argument ). The
argument is developed in detail in many texts (for example, de Finetti,
1974) and I will not repeat it here. Since at the outset the agent does not
know which measurement $\mathcal{B}\in \mathcal{M}$ will be chosen by the
bookie RULE 1 follows.

\textbf{RULE 2:} \textit{If }$\mathcal{B}_{1},\mathcal{B}_{2}\in \mathcal{M}$%
,$\ F\in \mathcal{B}_{1}\cap \mathcal{B}_{2}$\textit{\ then }$p(F\mid 
\mathcal{B}_{1})=p(F\mid \mathcal{B}_{2})$.

The rule asserts the non-contextuality of probability (Barnum et al, 2000).
It is not so much a rule of rationality, rather it is related to the logic
of the gamble and the identity of observables (remembering that we identify
each observable with the Boolean algebra generated by its possible outcomes).

Suppose that in the game $\mathcal{M}$, there are two measurements $\mathcal{%
B}_{1},\mathcal{B}_{2}\in \mathcal{M}$, and an event $F\in \mathcal{B}%
_{1}\cap \mathcal{B}_{2}$. Assume that an agent chooses to assign \textit{\ }%
$p(F\mid \mathcal{B}_{1})\neq p(F\mid \mathcal{B}_{2})$. A natural question
to ask her then is why she assigns $F$ different probabilities in the two
contexts, though she thinks it is the same event. The only answer consistent
with Bayesian probability theory is that she takes the $p(F\mid \mathcal{B}%
_{i})$ as \textit{conditional probabilities }and therefore not necessarily
equal. In other words, she considers the act of choosing an experiment $%
\mathcal{B}_{i}$ (in stage 3 of the gamble) as an event in a larger algebra $%
\mathcal{B}$ which contains $\mathcal{B}_{1},\mathcal{B}_{2}$. Consequently
she calculates the conditional probability of $F$, given the choice of $%
\mathcal{B}_{i}$.

There are two problems with this view. Firstly, the agent can no longer
maintain that $F\in \mathcal{B}_{1}\cap \mathcal{B}_{2}$, in fact $F$ is not
an element of any of the $\mathcal{B}_{i}$'s and can no longer be described
as \textit{an outcome }of a mesurement. Secondly, the agent assumes that
there is a single ``big'' Boolean algebra $\mathcal{B}$, the event $F$ is an
element of $\mathcal{B}$, and $\mathcal{B}_{1},\mathcal{B}_{2}$ are
sub-algebras of $\mathcal{B}$. The trouble is that\textit{\ for sufficiently
rich games} $\mathcal{M}$\textit{\ this assumption is inconsistent. }In
other words, there are gambles $\mathcal{M}$ which cannot be imbedded in a
Boolean algebra without destroying the identities of the events and the
logical relations between them. This is a consequence of the Kochen and
Specker (1967) theorem to which we shall come in (2.2). It means,
essentially, that an agent who violates RULE\ 2, is failing to grasp the
logic of the gamble and wrongly assumes that she is playing a different game.

Another possibility is that assigning $p(F\mid \mathcal{B}_{1})\neq p(F\mid 
\mathcal{B}_{2})$ indicates that the agent is using a different notion of
conditional probability. The burden of clarification is then on the agent,
to uncover her sense of conditionalization and show how it is related to
quantum gambles. Thus, we conclude that the violation of RULE 2 implies
either an ignorance of the logic of the gamble, or an incoherent use of
conditional probabilities. It is clear that our argument here is weaker than
the Dutch book argument for RULE 1. A violation of RULE 2 does not imply a
sure loss in a single shot game. We shall return to this argument, with a
greater detail in section 2.2.

Rational probability values assigned in finite games need not be numerically
identical to the quantum mechanical probabilities. However, with
sufficiently complex gambles we can show that all the interesting features
of quantum probability- from the uncertainty principle to the violation of
Bell inequality-are present even in finite gambles. If we extend our
discussion to gambles with an infinity of possible measurements, then RULE1
and RULE 2 force the probabilities to follow Born rule (section 2.4).

\subsection{A Note on Possible Games}

A quantum gamble is a set of Boolean algebras with certain (possible)
relations between them. The details of these algebras and their relations is
all that the agent needs to know. We do not assume that the agent knows any
quantum theory.

However, engineers who construct gambling devices should know a little more.
They should be aware of the physical possibilities. This is true in the
classical domain as much as in the quantum domain. After all, the theory of
probability, even in its most subjective form, associates a person's degree
of belief with the objective possibilities in the physical world. In the
quantum case the objective physical part concerns the \textit{type} of
gambles which can actually be constructed. It turns out that not all finite
families of Boolean algebras represent possible games, at least as far as
present day physics is concerned. I shall describe the family of possible
gambles, in a somewhat abstract way. It is a consequence of von Neumann
(1955) analysis of the set of possible measurements.

Let $\mathbb{H}$ be the $n$-dimensional vector space over the real or
complex field, equipped with the usual inner product. Let $%
H_{1},H_{2},...,H_{k}$ be $k$ non zero subspaces of $\mathbb{H}$, which are
orthogonal in pairs $H_{i}\perp H_{j}$ for $i,j=1,2,...,k$, and which
together span the entire space, $H_{1}\oplus H_{2}\oplus ...\oplus H_{k}=%
\mathbb{H}$. These subspaces generate a Boolean algebra, call it $\mathcal{B}%
(H_{1},H_{2},...,H_{k})$, in the following way: The zero of the algebra is
the null subspace, the non zero elements of the algebra are subspaces the
form $H_{i_{1}}\oplus H_{i_{2}}\oplus ...\oplus H_{i_{r}}$ where $\phi \neq
\{i_{1},i_{2,}...,i_{r}\}\subseteq \{1,2,...,k\}$. If $H,H`$ are two
elements in the algebra let $H\vee H`=H\oplus H`$ be the subspace spanned by
the (set theoretic) union $H\cup H`$, let $H\wedge H`=H\cap H`$, and let the
complement of $H$, denoted by $H^{\perp }$, be the subspace orthogonal to $H$
such that $H\oplus H^{\perp }=\mathbb{H}$. Then $\mathcal{B}%
(H_{1},H_{2},...,H_{k})$ with the operations $\vee $, $\wedge $, $\perp $ is
a Boolean algebra with $2^{k}$ elements. Note that a maximal algebra of this
kind is obtained when we take all the $H_{i}$`s to be one-dimensional
subspaces (rays). Then $k=n$ and the algebra has $2^{n}$ elements.

Now, let $\mathbb{B(H})$ be the family of \textit{all} the Boolean algebras
obtained from subspaces of $\mathbb{H}$\ in the way described above.
Obviously, If $\mathcal{B}_{1},\mathcal{B}_{2}\in \mathbb{B(H})$ then $%
\mathcal{B}_{1}\cap \mathcal{B}_{2}$ is also Boolean algebra in $\mathbb{B(H}%
)$. We shall say that two subspaces\textit{\ }$G,H$\textit{\ }of\textit{\ }$%
\mathbb{H}$\textit{\ }are\textit{\ compatible in }$\mathbb{H}$\textit{\ }if
there is\textit{\ }$\mathcal{B}\in \mathbb{B}(\mathbb{H})$\textit{\ }such
that\textit{\ }$G,H\in \mathcal{B}$\textit{, }otherwise $G$\textit{\ }and%
\textit{\ }$H$\textit{\ }are incompatible. Two algebras $\mathcal{B}_{1},%
\mathcal{B}_{2}$ are \textit{incompatible in }$\mathbb{H}$ if there are
subspaces $G\in \mathcal{B}_{1}$ and $H\in \mathcal{B}_{2}$ which are
incompatible.

\textbf{POSSIBILITY CRITERION:} $\mathcal{M}$ \textit{is a possible quantum
gamble if there is a finite dimensional complex or real Hilbert space} $%
\mathbb{H}$ \textit{such that} $\mathcal{M}$ \textit{is a finite family of
Boolean algebras in} $\mathbb{B(H})$ \textit{which are incompatible in pairs}%
.

One could proceed with the probabilistic account disregarding this criterion
and, in fact, go beyond what is known to be physically possible (see Svozil,
1998). We shall not do that, however, and all the games considered in this
paper are physically possible. With each of the gambles to be discussed in
this paper we proceed in two stages. Firstly, we present the Boolean
algebras, their relations and the consequences for probability. Secondly, we
prove that the gamble obeys the possibility criterion.

\section{Consequences}

\subsection{Uncertainty Relations}

Consider the following quantum gamble $\mathcal{M}$ consisting of seven
incompatible measurements (Boolean algebras), each generated by its three
possible outcomes: $\left\langle E_{1},E_{2},F_{2}\right\rangle $, $%
\left\langle E_{1},E_{3},F_{3}\right\rangle $, $\left\langle
E_{2},E_{4},E_{6}\right\rangle $, $\left\langle
E_{3},E_{5},E_{7}\right\rangle $, $\left\langle E_{6},E_{7},F\right\rangle $%
, $\left\langle E_{4},E_{8},F_{4}\right\rangle $, $\left\langle
E_{5},E_{8},F_{5}\right\rangle $. Note that some of the outcomes are shared
by two measurements, these are denoted by the letter $E$. The other outcomes
belong each to a single algebra and denoted by $F$. As before, when two
algebras share an event they also share its complement so that, for example,$%
\overline{E}_{1}=E_{2}\cup F_{2}=E_{3}\cup F_{3}$, and similarly in the
other cases. The logical relations among the generators are depicted in the
graph of figure 2. This is \textit{the compatibility graph} of the
generators. Each node in the graph represents an outcome, two nodes are
connected by an edge if, and only if the corresponding outcomes belong to a
common algebra; each triangle represents the generators of one of the
algebras.

\begin{center}
\FRAME{ftbpF}{2.879in}{2.1612in}{0pt}{}{}{bayes02.eps}{\special{language
"Scientific Word";type "GRAPHIC";maintain-aspect-ratio TRUE;display
"USEDEF";valid_file "F";width 2.879in;height 2.1612in;depth
0pt;original-width 8.534in;original-height 6.3866in;cropleft "0";croptop
"1";cropright "1";cropbottom "0";filename 'Bayes02.eps';file-properties
"XNPEU";}}
\end{center}

We assume that the agent is aware of the seven algebras and the connections
between them. By RULE 2 the probability he assigns to each event is
independent of the Boolean algebra (measurement) which is considered, for
example, $p(E_{2}\left| \left\langle E_{1},E_{2},F_{2}\right\rangle \right.
)=p(E_{2}\left| \left\langle E_{2},E_{4},E_{6}\right\rangle \right. )\equiv
p(E_{2})$. RULE 1 entails that the probabilities of each triple of outcomes
of each measurement should sum up to 1, for example, $%
p(E_{4})+p(E_{8})+p(F_{4})=1$. There are altogether seven equations of this
kind. Combining them with the fact that probability is non-negative (by
RULE\ 1) it is easy to prove that the probabilities assigned by our rational
agent should satisfy $p(E_{1})+p(E_{8})\leq \frac{3}{2}$. This is an example
of an \textit{uncertainty relation}, a constraint on the probabilities
assigned to the outcomes of incompatible measurements. In particular, if the
system is prepared in such a way that it is rational to assign $p(E_{1})=1$
(see 2.5) then the rules of quantum games force $p(E_{8})\leq \frac{1}{2}$.

To see why $\mathcal{M}$ represents a physically possible gamble we use the
POSSIBILITY CRITERION\ and identify each event with a one dimensional
subspace of \ $\mathbb{C}^{3}$ (or $\mathbb{R}^{3}$) in the following way $%
E_{1}$ is the subspace spanned by the vector $(1,0,2)$, $E_{2}\backsim
(0,1,0)$, $F_{2}\backsim (2,0,-1)$, $E_{3}\backsim (2,1,-1)$, $F_{3}\backsim
(2,-5,-1)$, $E_{4}\backsim (0,0,1)$, $E_{5}\backsim (1,-1,1)$, $%
E_{6}=(1,0,0) $, $E_{7}\backsim (0,1,1)$, $F\backsim (0,1,-1)$, $%
F_{4}\backsim (1,-1,0)$, $F_{5}\backsim (-1,1,2)$, $E_{8}\backsim (1,1,0)$.
Note that the vectors associated with compatible subspaces are orthogonal,
so that figure 2 is the orthogonality graph for these thirteen vectors.

A more concrete way to represent this game is to consider each of these
vectors as depicting a direction in physical space. For the vector $v$ let $%
S_{v}^{2}$ be the square of the spin in the $v$-direction of a massive
spin-1 particle, so that its eigenvalues are $0,1$. Now, for each of the
thirteen vectors above take the event $\{S_{v}^{2}=0\}$. Then the relations
in figure 2 are satisfied.

This example is a special case of a more general principle (Pitowsky, 1998):

\begin{theorem}
let $H_{1},H_{2}$ \textit{be two incompatible rays in a Hilbert space }$%
\mathbb{H}$ \ whose dimension $\geq 3$. Then there is a (finite) quantum
gamble $\mathcal{M}\subset \mathbb{B(H})$ in which $H_{1},H_{2}$ are events,
and every probability assignment $p$ for $\mathcal{M}$ which satisfies RULE
1 and RULE 2 also satisfies $p(H_{1})+p(H_{2})<1$.
\end{theorem}

\subsection{Truth and Probability, The Kochen and Specker's Theorem}

Consider the gamble $\mathcal{M}$ of eleven incompatible measurements, each
with four possible outcomes.

$\mathcal{B}_{1}=\left\langle E_{1},F_{1},F_{2},F_{3}\right\rangle $, $%
\mathcal{B}_{2}=\left\langle E_{1},F_{1},F_{4},F_{5}\right\rangle $, $%
\mathcal{B}_{3}=\left\langle E_{1},F_{2},F_{6},F_{7}\right\rangle $,

$\mathcal{B}_{4}=\left\langle E_{1},F_{3},F_{8},F_{9}\right\rangle $, $%
\mathcal{B}_{5}=\left\langle E_{2},F_{10},F_{11},F_{12}\right\rangle $, $%
\mathcal{B}_{6}=\left\langle E_{2},F_{7},F_{10},F_{13}\right\rangle $,

$\mathcal{B}_{7}=\left\langle E_{2},F_{8},F_{11},F_{14}\right\rangle $, $%
\mathcal{B}_{8}=\left\langle E_{2},F_{4},F_{12},F_{15}\right\rangle $, $%
\mathcal{B}_{9}=\left\langle F_{9},F_{14},F_{16},F_{17}\right\rangle $,

$\mathcal{B}_{10}=\left\langle F_{5},F_{15},F_{16},F_{18}\right\rangle $, $%
\mathcal{B}_{11}=\left\langle F_{6},F_{12},F_{17},F_{18}\right\rangle $

The two outcomes denoted by the letter $E$ are shared by four measurement
each, and the outcomes denoted by $F$ are shared by two measurements each.
Altogether there are twenty outcomes. This example is based on a proof of
the Kochen and Specker (1967) theorem due to Kargnahan(1994). (The original
proof requires hundreds of measurements, with three outcomes each and 117
outcomes in all). Again, when an event is shared by two measurements then so
does its complement, for example, $\overline{F}_{8}=E_{1}\cup F_{3}\cup
F_{9}=E_{2}\cup F_{11}\cup F_{14}$.

Now, suppose that all the algebras $\mathcal{B}_{k}$ are sub-algebras of a
Boolean algebra $\mathcal{B}$. Assume, without loss of generality, that $%
\mathcal{B}$ is an algebra of subsets of a set $X$. With this identification
the events $E_{i},F_{j}$ are subsets of $X$. The logical relations between
the events dictates that any two of the events among the $E_{i}$'s and $%
F_{j} $'s that share the same algebra $\mathcal{B}_{k}$ are disjoint.
Moreover, the union of all four outcomes in each algebra $\mathcal{B}_{k}$,
is identical to $X$, for example, $X=E_{2}\cup F_{7}\cup F_{10}\cup F_{13}$
is the union of the outcomes in $\mathcal{B}_{6}$. But this leads to a
contradiction because the intersection of all these unions is necessarily
empty!

To see that suppose, by contrast, that there is $x\in X$ such that $x$
belongs to exactly one outcome, $E_{i}$ or $F_{j}$, in each one of the
eleven algebras $\mathcal{B}_{k}$. This means that $x$ belong to eleven such
events (with repetition counted). But this is impossible since each one of
the outcomes appears an even number of times in the eleven algebras, and
eleven is an odd number.

One consequence of this is related to RULE\ 2 discussed in section 1.3.
Suppose that an agent thinks about the probabilities of the events $%
E_{i},F_{j}$ as conditional on the measurement performed. If the term
``conditional probability'' is used in its usual sense then the events
should be interpreted as elements of a single Boolean algebra $\mathcal{B}$
(taken again as an algebra of subsets of some set $X$). To avoid the Kochen
Specker contradiction the agent can use two strategies. The first to take
some of the generating events in at least one algebra to be non-disjoint in
pairs,\ for example, $E_{2}\cap F_{8}\neq \phi $. In this case the agent
seizes to see the events $E_{2},F_{8}$ as representing \textit{measurement
outcomes,} and associates with them some other meaning (although he
eventually takes the conditional probability of $E_{2}\cap F_{8}$ to be
zero). The other strategy is to take the union of \ the outcomes of some
measurements to be proper subset of $X$. For example, in the case of $%
\mathcal{B}_{9}$, $F_{9}\cup F_{14}\cup F_{16}\cup F_{17}\varsubsetneq X$.
In this case the agent actually adds another theoretical outcome (which,
however, has conditional probability zero). Both strategies represent a
distortion of the logical relations among the events, which we have assumed
as given.

On a less formal level we can ask why would anyone do that? The additional
structure assumed by the agent amounts to a strange ``hidden variable
theory'' for the set of experiments $\mathcal{M}$. There is a great
theoretical interest in hidden variable theories, but they are of little
value to the rational gambler. A classical analogue would be a person who
thinks that a coin \textit{really} has three sides `head', `belly' and
`tail' and assigns a prior probability $\frac{1}{3}$ to each. But the act of
tossing the coin (or looking at it, or physically interacting with it)
causes the belly side never to show up, so the probability of belly,
conditional on tossing (or looking, or interacting), is zero. The betting
behavior of such a person is rational in the sense that no Dutch book
argument against him is possible. However, as far as gambling on a coin toss
is concerned, his theory of coins is not altogether rational. It is the
elimination of this kind of irrationality which motivates RULE 2.

Another consequence of this gamble concerns the relations between
probability and logical truth. Often the Kochen and Specker theorem is taken
as an indication that in quantum mechanics a classical logical falsity may
sometimes be true (Bub, 1974; Demopoulos, 1976). To see how, consider the $%
E_{i}$ and $F_{j}$ as \textit{propositional variables,} and for each $1\leq
k\leq 11$ let $C_{k}$ be the proposition which says: ``exactly one of the
variable in the group $k$ is true'', for example,

\begin{eqnarray*}
C_{6} &=&(E_{2}\vee F_{7}\vee F_{10}\vee F_{13})\wedge \thicksim
(E_{2}\wedge F_{7})\wedge \thicksim (E_{2}\wedge F_{10})\wedge \smallskip \\
&\thicksim &(E_{2}\wedge F_{13})\wedge \thicksim (F_{7}\wedge F_{10})\wedge
\thicksim (F_{7}\wedge F_{13})\wedge \thicksim (F_{10}\wedge F_{13})
\end{eqnarray*}
Then $\bigwedge_{k=1}^{11}C_{k}$ is a classical logical falsity. But $%
\bigwedge_{k=1}^{11}C_{k}$ is `quantum mechanically true' with respect to
the system described above, because each one of the $C_{k}$'s is a true
description of it.

In our gambling picture we make a more modest claim. A rational agent who
participates in the quantum gamble will assign, in advance, probability $1$
to each $C_{k}$. Therefore, arguably the agent also assigns $%
\bigwedge_{k=1}^{11}C_{k}$ probability $1$. But this is an \textit{epistemic}
position which does not oblige the agent to assign truth values to the $%
E_{i} $'s and $F_{j}$'s, nor is he committed to say that such truth values
exist. Indeed, this is a strong indication that `probability one' and
`truth' are quite different from one another. The EPR system (below)
provides another example. There is, however, a weaker sense in which $%
\bigwedge_{k=1}^{11}C_{k}$ is true and we shall discuss it in the
philosophical discussion 3.1.

The following is a proof that our game satisfies the POSSIBILITY CRITERION.
Each $E_{i}$ and each $F_{j}$ is identified with a ray (one dimensional
subspace) of \ $\mathbb{C}^{4}$ (or $\mathbb{R}^{4}$). Two outcome which
share the same algebra correspond to orthogonal rays. The rays are
identified by a vector that spans them:

$E_{1}\backsim (1,0,0,0)$, $F_{1}\backsim (0,1,0,0)$, $F_{2}\backsim
(0,0,1,0)$, $F_{3}\backsim (0,0,0,1)$,

$F_{4}\backsim (0,0,1,1)$, $F_{5}\backsim (0,0,1,-1)$, $F_{6}\backsim
(0,1,0,1)$, $F_{7}\backsim (0,1,0,-1)$,

$F_{8}\backsim (0,1,1,0)$, $F_{9}\backsim (0,1,-1,0)$, $E_{2}\backsim
(1,1,-1,1)$, $F_{10}\backsim (-1,1,1,1)$,

$F_{11}\backsim (1,-1,1,1)$, $F_{12}\backsim (1,1,1,-1)$, $F_{13}\backsim
(1,0,1,0)$, $F_{14}\backsim (1,0,0,-1)$,

$F_{15}\backsim (1,-1,0,0)$, $F_{16}\backsim (1,1,1,1)$, $F_{17}\backsim
(1,-1,-1,1)$, $F_{18}\backsim (1,1,-1,-1)$.

\subsection{EPR\ and Violation of Bell's Inequality}

Given two (not necessarily disjoint) events $A$, $B$ in the same algebra,
denote $AB=A\cap B$, and for three events $A,B,C$ denote by $\{A,B,C\}$ the
Boolean algebra that they generate:

\begin{equation*}
\{A,B,C\}=\left\langle ABC,\ \overline{A}BC,\ A\overline{B}C,\ AB\overline{C}%
,\ \overline{A}\overline{B}C,\ \overline{A}B\overline{C},\ A\overline{B}%
\overline{C},\ \overline{A}\overline{B}\overline{C}\right\rangle
\end{equation*}

In order to recover the argument of the Einstein Rosen and Podolsky (1935)
and Bell (1966) paradox within a\ quantum gamble we shall use Mermin (1990)
representation of GHZ, the Greenberger-Horne-Zeilinger (1989)\ system.
Consider the gamble which consists of eight possible measurements: The four
measurements $\mathcal{B}_{1}=\{A_{1},B_{1},C_{1}\}$, $\mathcal{B}%
_{2}=\{A_{1},B_{2},C_{2}\}$, $\mathcal{B}_{3}=\{A_{2},B_{1},C_{2}\}$, $%
\mathcal{B}_{4}=\{A_{2},B_{2},C_{1}\}$ each with eight possible outcomes and

$\mathcal{B}_{5}=\left\langle S,\ D_{1},\ A_{1}B_{1}C_{1},\ \overline{A}_{1}%
\overline{B}_{1}C_{1},\ \overline{A}_{1}B_{1}\overline{C}_{1},\ A_{1}%
\overline{B}_{1}\overline{C}_{1}\right\rangle $,

$\mathcal{B}_{6}=\left\langle S,\ D_{2},\ \overline{A}_{1}B_{2}C_{2},\ A_{1}%
\overline{B}_{2}C_{2},\ A_{1}B_{2}\overline{C}_{2},\ \overline{A}_{1}%
\overline{B}_{2}\overline{C}_{2}\right\rangle $,

$\mathcal{B}_{7}=\left\langle S,\ D_{3},\ \overline{A}_{2}B_{1}C_{2},\ A_{2}%
\overline{B}_{1}C_{2},\ A_{2}B_{1}\overline{C}_{2},\ \overline{A}_{2}%
\overline{B}_{1}\overline{C}_{2}\right\rangle $,

$\mathcal{B}_{8}=\left\langle S,\ D_{4},\ \overline{A}_{2}B_{2}C_{1},\ A_{2}%
\overline{B}_{2}C_{1},\ A_{2}B_{2}\overline{C}_{1},\ \overline{A}_{2}%
\overline{B}_{2}\overline{C}_{1}\right\rangle $,

\noindent each with six possible outcomes.

Assume that the agent has good reasons to believe that $p(S)=1$. Such a
belief can come about in a variety of ways, for example, she may know
something about the preparation of the system form a previous measurement
result (see section 2.5). Alternatively, the bookie may announce in advance
that he will raise his stakes indefinitely against any bet made for $%
\overline{S}$. Whatever the source of information, the agent has good
reasons to assign probability zero to four out of the eight outcomes in each
one of the four measurements $\mathcal{B}_{1}$ to $\mathcal{B}_{4}$. The
remaining events are

\begin{eqnarray}
&&in\ \mathcal{B}_{1}\quad \overline{A}_{1}B_{1}C_{1},\ A_{1}\overline{B}%
_{1}C_{1},\ A_{1}B_{1}\overline{C}_{1},\ \overline{A}_{1}\overline{B}_{1}%
\overline{C}_{1}\smallskip \\
&&in\ \mathcal{B}_{2}\quad A_{1}B_{2}C_{2},\ \overline{A}_{1}\overline{B}%
_{2}C_{2},\ \overline{A}_{1}B_{2}\overline{C}_{2},\ A_{1}\overline{B}_{2}%
\overline{C}_{2}\smallskip  \notag \\
&&in\ \mathcal{B}_{3}\quad A_{2}B_{1}C_{2},\ \overline{A}_{2}\overline{B}%
_{1}C_{2},\ \overline{A}_{2}B_{1}\overline{C}_{2},\ A_{2}\overline{B}_{1}%
\overline{C}_{2}\smallskip  \notag \\
&&in\ \mathcal{B}_{4}\quad A_{2}B_{2}C_{1},\ \overline{A}_{2}\overline{B}%
_{2}C_{1},\ \overline{A}_{2}B_{2}\overline{C}_{1},\ A_{2}\overline{B}_{2}%
\overline{C}_{1}  \notag
\end{eqnarray}
Denote by $P$ the sum of the probabilities of these sixteen events. Given
that $p(S)=1$ the probabilities of the events in each row in (1)\textit{\ }%
sum up to $1$. Altogether, the rational assignment is therefore $P=4$.
However, \textit{if }$A_{1}$, $B_{1}$, $C_{1}$, $A_{2}$, $B_{2}$, $C_{2}$%
\textit{\ are events in any (classical) probability space then the sum of
the probabilities of the events in }(1)\textit{\ never exceeds }$3$. This is
one of the constraints on the values of probabilities which Boole called
``conditions of possible experience''\footnote{%
See Pitowsky (1989, 1994, 2002) and Pitowsky and Svozil, (2001) for a
discussion of Boole's conditions, their derivations and their violations by
quantum frequencies.} and it is violated by any rational assignment in this
quantum gamble. On one level this is just another example of a quantum
gamble that cannot be imbedded in a single classical probability space
without distorting the identity of the events and the logical relations
between them. A more dramatic example has been the Kochen and Specker's
theorem of the previous section.

The special importance of the EPR case lies in the details of the physical
system and the way the measurements $\mathcal{B}_{1},\mathcal{B}_{2},%
\mathcal{B}_{3},\mathcal{B}_{4}$ are performed. The system is composed of
three particles which interacted in the past, but are now spatially
separated and are no longer interacting. On the first particle we can choose
to perform an $A_{1}$-measurement or an $A_{2}$-measurement (but not both)
each with two possible outcomes. Similarly, we can choose to perform on the
second particle one of two $B$-measurement, and one of two $C$-measurement
on the third particle. The algebras $\mathcal{B}_{1},\mathcal{B}_{2},%
\mathcal{B}_{3},\mathcal{B}_{4}$ represent the outcomes of four out of the
eight logically possible combinations of such local measurements. In this
physical arrangement we can recover the EPR reasoning, and Bell's rebuttal,
which I will not repeat here. The essence of Bell's theorem is that the EPR
assumptions lead to the conclusion that $A_{1}$, $B_{1}$, $C_{1}$, $A_{2}$, $%
B_{2}$, $C_{2}$ belong to a single Boolean algebra. Consequently, the sum of
the probabilities of the events in (1) cannot exceed $3$, in contradiction
to RULE1 and RULE 2.

Which of two EPR assumptions `reality' or `locality' should the Bayesian
reject? In the previous section we have made the distinction between
`probability $1$' and `truth'. But the \textit{identification} of the two is
precisely the subject matter of EPR's Principle of Reality: ``If without in
any way disturbing a system we can predict with certainty (i.e. with
probability equal to unity) the value of a physical quantity, then there
exists an element of reality corresponding to this physical quantity''
(Einstein Rosen and Podolsky, 1935)\textit{.} Quite independently of Bell's
argument, a Bayesian should take a sceptical view of this principle.
``Probability equal to unity'' means that the degree of rational belief has
reached a level of certainty. It does not reflect any prejudice about
possible causes of the outcomes. On the other hand, there seem to be no good
grounds for rejecting the Principle of Locality on the basis of this or
similar gambles.

To prove that this gamble satisfies the possibility criterion let $\mathbb{H}%
_{2}$ be the two dimensional complex Hilbert space, let $\sigma _{x}$, $%
\sigma _{y}$ be the Pauli matrices associated with the two orthogonal
directions $x$, $y$, and let $H_{x}$, $H_{y}$ the (one dimensional)
subspaces of $\mathbb{H}_{2}$ corresponding to the eigenvalues $\sigma
_{x}=1 $, $\sigma _{y}=1$ respectively, so that $H_{x}^{\bot }$, $%
H_{y}^{\bot }$ correspond to $\sigma _{x}=-1$, $\sigma _{y}=-1$. In the
eight dimensional Hilbert space $\mathbb{H}_{2}\otimes \mathbb{H}_{2}\otimes 
\mathbb{H}_{2}$ we shall identify $A_{1}=H_{x}\otimes \mathbb{H}_{2}\otimes 
\mathbb{H}_{2}$, $B_{1}=\mathbb{H}_{2}\otimes H_{x}\otimes \mathbb{H}_{2}$, $%
C_{1}=\mathbb{H}_{2}\otimes \mathbb{H}_{2}\otimes H_{x}$, $%
A_{2}=H_{y}\otimes \mathbb{H}_{2}\otimes \mathbb{H}_{2}$, $B_{2}=\mathbb{H}%
_{2}\otimes H_{y}\otimes \mathbb{H}_{2}$, $C_{2}=\mathbb{H}_{2}\otimes 
\mathbb{H}_{2}\otimes H_{y}$, all these are four dimensional subspaces. The
outcomes in $\mathcal{B}_{1},\mathcal{B}_{2},\mathcal{B}_{3},\mathcal{B}_{4}$
are one dimensional subspaces, for example $\overline{A}_{1}\overline{B}%
_{2}C_{2}=H_{x}^{\bot }\otimes H_{y}^{\bot }\otimes H_{y}$. The subspace $S$
is the one dimensional ray along the GHZ state $\sqrt{1/2}\left(
|+_{z}\rangle _{1}|+_{z}\rangle _{2}|+_{z}\rangle _{3}-|-_{z}\rangle
_{1}|-_{z}\rangle _{2}|-_{z}\rangle _{3}\right) $ where $z$ is the direction
orthogonal to $x$ and $y$. The subspaces $D_{i}$ are just the
orthocomplements, in $\mathbb{H}_{2}\otimes \mathbb{H}_{2}\otimes \mathbb{H}%
_{2}$, to the direct sum of the other subspaces in their respective
algebras. Hence, $\dim D_{i}=3$.

\subsection{The Infinite Gamble: Gleason's Theorem}

Let us take the idealization a step further. Assume that the bookie
announces that $\mathcal{M}$ contains all the maximal Boolean algebras in $%
\mathbb{B(H})$ for some finite dimensional real or complex Hilbert space $%
\mathbb{H}$ with $\dim \mathbb{H}=n\mathbb{\geq }3$. Recall that if $%
H_{1},H_{2},...,H_{k}$ are $k$ non zero subspaces of $\mathbb{H}$, which are
orthogonal in pairs, and whose direct sum is the entire space, they generate
a Boolean algebra $\mathcal{B}(H_{1},H_{2},...,H_{k})$ (section 1.4). If $%
k=n $ the algebra is maximal and each subspace $H_{j}$ is one dimensional.
In other words, the set $\mathcal{M}$ consists of all non-degenerate
measurements with $n$ outcomes. The information theoretic aspects of this
case are discussed in Caves, Fuchs and Schack (2002)

There is a certain difficulty in extending quantum gambles to this case
since there are a continuum of possible measurements, and the agent is
supposed to place money on each. We can overcome this difficulty by assuming
that the agent makes a commitment to pay a certain amount on each outcome of
each measurement, without paying any cash in advance. When a single
measurement $\mathcal{B}\in \mathcal{M}$ is chosen by the bookie all the
agent's commitments are canceled, except those pertaining to $\mathcal{B}$.

RULE 1 and RULE 2 imply in this case that for any $n$ orthogonal rays $%
H_{1},H_{2},...,H_{n}$ in $\mathbb{H}$ the agent's probability function
should satisfy

\begin{equation}
p(H_{1})+p(H_{2})+...+p(H_{n})=1
\end{equation}

Gleason (1957) proved

\begin{theorem}
Let $\mathbb{H}$ be a Hilbert space over field of real or complex numbers
with a finite dimension $n\geq 3$. If $p$ is a non negative function defined
on the subspaces of $\mathbb{H}$ $\ $and satisfies (2) for every set of $n$
orthogonal rays then there is a statistical operator $W$ such that for every
subspace $H$ of $\mathbb{H}$ 
\begin{equation}
p(H)=tr(WP_{H})
\end{equation}
where $P_{H}$ is the projection operator on $H$.
\end{theorem}

For the proof see also Pitowsky (1998). This profound theorem gives a
characterization of all probability assignments of quantum theory.
Furthermore, if we know that the system is prepared with $p(R)=1$, for some
ray $R$, then $p$ is uniquely determined by $p(H)=\left\| P_{H}(r)\right\|
^{2}$ for all subspaces $H$, where $r$ is a unit vector that spans $R$. The
theorem can be easily extended to closed subspaces of the infinite
dimensional Hilbert space.

It is interesting to note that many of the results about finite quantum
gambles that we have considered are actually consequences of Gleason's
theorem. Consider, for example the Kochen and Specker's theorem (section
2.2). To connect it with Gleason's theorem take an appropriate first order
formal theory of the rays of $\mathbb{R}^{n}$, the orthogonality relation
between them, and the real functions defined on them (where $n\geq 3$ finite
and fixed). Add to it a special function symbol $p$, the axiom that $p$ is
non negative, the axiom that $p$ is not a constant, the axiom that $p$ has
only two values zero or one. Now, add the infinitely many axioms $%
p(H_{1})+p(H_{2})+...+p(H_{n})=1$ for each $n$-tuple of orthogonal rays in $%
\mathbb{R}^{n}$. By Gleason's theorem this theory is inconsistent (since by
(3) $p$ has a continuum of values). Hence, there is a finite subset of this
set of axioms which is inconsistent, meaning a finite subset of rays which
satisfy the Kochen and Specker's theorem. This is, of course, a non
constructive proof, and an explicit construction is preferable. However, the
consideration just mentioned can be used to obtain more general
non-constructive results about finite games. One such immediate result is
Theorem 1 which also has a constructive proof. (In fact, the proof of
Gleason's theorem involves a construction similar to that of theorem 1, see
Pitowsky, 1998)

Gleason's theorem indicates that the use of the adjective `subjective' to
describe epistemic probability is a misnomer. Even in the classical realm it
has misleading connotations. Classically, different agents that start with
different prior probability assignments eventually converge on the same
probability distribution as they learn more and more from common experience.
In the quantum realm the situation is more extreme. For a given a single
physical system Gleason's theorem dictates that all agents share a common
prior or, in the worst case, they start using the same probability
distribution after a single (maximal) measurement.

\subsection{A Note on Conditional Quantum Probability}

Consider two gambles, $\mathcal{M}_{1}$, $\mathcal{M}_{2}$ and assume that $%
A $ is a common event. In other words, there is $\mathcal{B}_{1}\in \mathcal{%
M}_{1}$ and $\mathcal{B}_{2}\in \mathcal{M}_{2}$ such that $A\in \mathcal{B}%
_{1}\cap \mathcal{B}_{2}$. We can consider sequential gambles in which the
gamble $\mathcal{M}_{1}$ is played, and subsequently after the results are
recorded, the gamble $\mathcal{M}_{2}$ follows with the measurements
performed \textit{on the same system. }In such cases the agent can place 
\textit{conditional bets} of the form\textit{:} ``If $A$ occurs in the first
gamble place such and such odds in the second gamble''. This means that the
of probabilities assigned in the second game $\mathcal{M}_{2}$ are
constrained by the condition $p(A)=1$ (in addition to the constraints
imposed by RULE 1 and RULE\ 2). The EPR gamble in 2.3 can be seen as such a
conditional game, when we consider the preparation process as a previous
gamble with an outcome $S$. In fact, all preparations (at least of pure
states) can be seen in that light.

If the gambles $\mathcal{M}_{1}$, $\mathcal{M}_{2}$ are infinite, and
contain all the maximal algebras in $\mathbb{B(H})$, Gleason's theorem
dictates the rule for conditional betting. In the second gamble the
probability is the square of the length of the projection on (the subspace
corresponding to) $A$. The conditional probability is therefore given by
L\"{u}ders rule (Bub, 1997).

\section{Philosophical Remarks}

\subsection{Bohr, Quantum Logic and Structural Realism}

The line we have taken has some affinity with Bohr's approach -or more
precisely, with the view often attributed to Bohr\footnote{%
See Beller (1999). Although Bohr kept changing his views and contradicted
himself on occations, it is useful to distill from his various
pronouncements a more or less coherent set. This is what philosophers mean
by ``Bohr's views''.}-in that we treat the outcomes of future measurements
as mere possibilities, and do not associate them with properties that exist
prior to the act of measurement. Bohr's position, however, has some other
features which are better avoided. Consider a spin-1 massive particle and
suppose that we measure $S_{z}$, its spin along the $z$-direction. Bohr
would say that in this circumstance attributing values to $S_{x}$ and $S_{y}$
is meaningless. But the equation $S_{x}^{2}+S_{y}^{2}+S_{z}^{2}=2I$ remains
valid then, as it is valid at all times. How can an expression which
contains meaningless (or valueless) terms be itself valid? Indeed,
non-commuting observables may satisfy algebraic equations, the Laws of
Nature often take such form. What is the status of such equations at the
time when only one component in them has been meaningfully assigned a value?
What is their status when no measurement has been performed? Quantum logic,
in some of its formulations, has been an attempt to answer this question 
\textit{realistically.}

It had began with the seminal work of Birkhoff and von-Neumann (1936). A
later modification was inspired by the work of Kochen and Specker (1967).
The realist interpretation of the quantum logical formalism is due to
Finkelstein (1962), Putnam (1968), Bub (1974), Demopoulos (1976). Consider,
for example, the gamble $\mathcal{B}_{1}=\left\langle
E_{1},E_{2},E_{3}\right\rangle $, $\mathcal{B}_{2}=\left\langle
E_{1},E_{4},E_{5}\right\rangle $ made of two incompatible measurements, with
one common outcome $E_{1}$ (figure 1). Let us loosely identify the outcomes $%
E_{i}$ with the propositions that describe them. The realist quantum
logician maintains that both $E_{1}\vee E_{2}\vee E_{3}$ and $E_{1}\vee
E_{4}\vee E_{5}$ are true, and therefore so is $A=(E_{1}\vee E_{2}\vee
E_{3})\wedge (E_{1}\vee E_{4}\vee E_{5})$. But only one of the measurements $%
\mathcal{B}_{1}$ or $\mathcal{B}_{2}$ can be conducted at one time. This
means that, generally, only three out of the five $E_{i}$'s can be
experimentally assigned a truth value (except in the case that $E_{1}$ turns
out to be true which makes the other four events false). This does not
prevent us from assigning \textit{hypothetical} truth values to the $E_{i}$%
's that make $A$ true. However, as we have seen in 2.2, the trouble begins
when we consider more complex gambles. To repeat, let $\mathcal{M}$ be the
gamble of 2.2, and for each $1\leq k\leq 11$ let $C_{k}$ be the proposition
which says: ``exactly one of the variables in the group $k$ is true'', for
example,

\begin{eqnarray*}
C_{6} &=&(E_{2}\vee F_{7}\vee F_{10}\vee F_{13})\wedge \thicksim
(E_{2}\wedge F_{7})\wedge \thicksim (E_{2}\wedge F_{10})\wedge \smallskip \\
&\thicksim &(E_{2}\wedge F_{13})\wedge \thicksim (F_{7}\wedge F_{10})\wedge
\thicksim (F_{7}\wedge F_{13})\wedge \thicksim (F_{10}\wedge F_{13})
\end{eqnarray*}
Then $B=\bigwedge_{k=1}^{11}C_{k}$ is a classical logical falsity. This
means that we cannot make $B$ true even by assigning hypothetical truth
values to the $E_{i}$'s and $F_{j}$`s.

Still, the quantum logician maintains that $B$ is true. Or, by analogy, that 
$S_{x}^{2}+S_{y}^{2}+S_{z}^{2}=2I$ is true for every orthogonal triple $x$, $%
y$, $z$ in physical space. This is the quantum logical solution of the
Bohrian dilemma and it comes with a heavy price-tag: the repudiation of
classical propositional logic. But what does it mean to say that $B$ is
true? As I have shown elsewhere (Pitowsky, 1989) the operational analysis of
the quantum logical connectives, due to Finkelstein and Putnam, leads to a
non-local hidden variable theory in disguise. Moreover, from a Bayesian
perspective it is quite sufficient to say that $B$ has probability $1$,
meaning that each conjunct in $B$ has probability $1$ that is, a degree of
belief approaching certainty. Indeed, the Bayesian does not consider even
the Laws of Nature as true, only as being nearly certain, given present day
knowledge.

Nevertheless, there is a sense in which $A$ or even $B$ are true, and this
is the sense that enables our Bayesian analysis in the first place. Thus, 
\textit{to assert that ``}$(E_{1}\vee E_{2}\vee E_{3})\wedge (E_{1}\vee
E_{4}\vee E_{5})$\textit{\ is true'' is nothing but a cumbersome way to say
that the gamble }$\mathcal{M}=\{\left\langle E_{1},E_{2},E_{3}\right\rangle
,\left\langle E_{1},E_{4},E_{5}\right\rangle \}$\textit{\ exists}. This is
first and foremost a statement about the identities: the outcome $E_{1}$ is 
\textit{really} the same in the two measurements, and $\overline{E}%
_{1}=E_{2}\vee E_{3}=E_{4}\vee E_{5}$. It is also a statement about physical
realizations, this gamble can be designed and played (experimental
difficulties notwithstanding). Viewed in this light quantum gambles together
with RULE 1 and RULE 2 form\textit{\ semantics for quantum logic,} in that
they assign meaning to the identities of quantum logic (in its partial
Boolean algebra formulation)\textit{.}

The metaphysical assumption underlying the Bayesian approach is therefore 
\textit{realism about the structure of quantum gambles}, in particular those
that satisfy the possibility criterion (1.4). This position is close in
spirit (but not identical) to the view that quantum mechanics is a complete
theory, so let us turn to the alternative view.

\subsection{Hidden Variables- A Bayesian Perspective}

Consider Bohm's theory as a typical example\footnote{%
The uniqueness theorem (Bub and Clifton 1996; Bub 1997; Bub Clifton and
Goldstein 2000) implies that all `no collapse'hidden variable theories have
essentially the structure of Bohm's theory.}. Recall that in this theory the
state of a single particle at time $t$ is given by the pair $(x(t),\psi
(x,t))$ where $x$ is the position of the particle and $\psi =R\exp (iS)$
-the guiding wave- is a solution of the time dependent Schr\"{o}dinger's
equation. The guiding condition $m\overset{\cdot }{x}=\nabla S$ provides the
relation between the two components of the state, where $m$ is the particle
mass. The theory is deterministic, an initial position $x(0)$ and an initial
condition $\psi (x,0)$ determine the trajectory of the particle and the
guiding wave at all future times. In particular, the outcome of every
measurement is determined by these initial conditions.

As can be expected from the Kochen and Specker's theorem the outcome of a
measurement is context dependent in Bohm's theory. This fact can also be
derived by a direct calculation (Pagonis and Clifton, 1995). Given a fixed
initial state $(x(0),\psi (x,0))$ the measurement of $S_{z}^{2}$ together
with $S_{x}^{2}$ and $S_{y}^{2}$ can yield one result $S_{z}^{2}=0$; but the
measurement of $S_{z}^{2}$ together with $S_{x`}^{2}$ and $S_{y`}^{2}$ can
give another result $S_{z}^{2}=1$. Now, the identity criterion for
observables in a deterministic theory is: \textit{Two procedures constitute
measurements of the same observable if for any given physical state
(preparation) they yield identical outcomes. }Therefore in Bohm's theory the
observable ``$S_{z}^{2}$ in the $x,y,z$ context'' is not really the same as
``$S_{z}^{2}$ in the $x`,$ $y`,$ $z$ context''. Nevertheless, the Bohmians
consider $S_{z}^{2}$ as one single \textit{statistical }observable across
contexts. The reason being that the \textit{average} outcome of $S_{z}^{2}$,
over different initial positions with density $\left| \psi (x,0)\right| ^{2}$%
, is context independent. Hence, Bohm's theory is a hybrid much like
classical statistical mechanics: the dynamics are deterministic but the
observables are statistical averages. Since the initial positions are not
known -not even knowable- the averages provide the empirical content.
Consequently, the observable structure of quantum mechanics is accepted by
the Bohmians ``for all practical purposes''.

This attitude prevails when the Bohmian is betting in a quantum gamble.
There is no detectable difference in the betting behavior of a Bohmian
agent; although the reasons leading to his behavior follow from the causal
structure of Bohm's theory. At a first glance there seems to be nothing
peculiar about this. Many people who would assign probability $0.5$ to
`heads' believe that the tossing of a coin is a deterministic process.
Indeed, there is a rational basis to this belief:\textit{\ }if the agent is
allowed to inspect the initial conditions of the toss with a greater
precision he may change his betting odds. In other words, his $0.5$ degree
of belief is \textit{conditional} on his lack of knowledge of the initial
state. Obtaining further information is possible, in principle, and in the
limit of infinite precision it leads to the assignment of probability zero
or one to `heads'.\ For the Bayesian this is in a large measure what
determinism \textit{means.}

Can we say the same about the Bohmian attitude in a quantum gamble?
According to Bohm's theory itself\footnote{%
Vallentini (1996) considers the possibility that $\left| \psi \right| ^{2}$
is only an `equilibrium' distribution, and deviations from it are possible.
In this case Bohm's theory is a genuine empirical extension of quantum
mechanics, and the Bohmian agent may sometime bet against the rules of
quantum mechanics.} the position of the particle cannot be known beyond the
information invested in the distribution $\left| \psi \right| ^{2}$. Suppose
that a particle is prepared in a (pure) quantum state $\psi (x,0)$. Then,
according to Bohm's theory, no further information is obtainable by a prior
inspection (without changing the quantum state, in which case the problem
starts all over again). Hence, $\left| \psi \right| ^{2}$ is an absolute,
not a conditional probability. Consequently, from a Bayesian perspective the
determinism of Bohm's theory is a myth. Luckily, it does not lead its
believers astray in their bets.

What is the function of this myth? Obviously, to retain a sense of
determinism, albeit one which is completely disconnected from human
knowledge. But there is also a subtler issue here that have to do with the
structure of the observables. As we have noticed, for the Bohmian the event $%
E_{1}=\{S_{z}^{2}=0$ in the $x,y,z$ context$\}$ is not the same as the event 
$E_{1}^{`}=\{S_{z}^{2}=0$ in the $x`,$ $y`,$ $z$ context$\}$. Hence, the
gamble $\mathcal{M}=\{\left\langle E_{1},E_{2},E_{3}\right\rangle $, $%
\left\langle E_{1},E_{4},E_{5}\right\rangle \}$ is interpreted by him as
being ``really'' $\mathcal{M}`=\{\left\langle E_{1},E_{2},E_{3}\right\rangle 
$, $\left\langle E_{1}^{`},E_{4},E_{5}\right\rangle \}$; although, as a
result of dynamical causes, the long term frequencies of $E_{1}$ and $%
E_{1}^{`}$ happen to be identical (for any given $\psi $). It follows that
the myth also serves the purpose of ``saving classical logic'' by dynamical
means (Pitowsky, 1994). Nowhere is this more apparent than in the EPR case
where Bohm's dynamics violate locality on the level of individual processes.

In this sense the hidden variable approach is conservative. It is not so
much its insistence on determinism, but rather the refusal to acknowledge
that the structure of the set of events- our quantum gambles- is real. As a
gambler the Bohmian bets as if it is very real; as a metaphysician he
provides a complicated apology.

\subsection{Instrumentalism and its Radical Foundations}

The Bayesian approach represents an instrumental attitude towards the
quantum state. The state is just a code for probabilities, and ``probability
theory is simply the quantitative formulation of how to make rational
decisions in the face of uncertainty'' (Fuchs and Peres, 2000).
Instrumentalism seems metaphysically innocent, all we are dealing with are
experiments and their outcomes, without a commitment to an underlying,
completely described microscopic reality. One might even be tempted to think
that ``quantum theory needs no interpretation'' (ibid). Of course, there is
a sense in which this is true. One needs no causal picture to do physics.
Like a gambler, the physicist can assign probabilities to outcomes, assuming
no causal or other mechanisms which bring them about.

But instrumentalism simply pushes the question of interpretation one step up
the ladder. Instead of dealing directly with `reality', the instrumentalist
faces the challenge of explaining his instrument, that is, quantum
probability. Unlike other mathematical theories- group theory for example-
the application of probability requires a philosophical analysis. After all
probability theory is our tool for weighing the relative merits of
alternative actions and for making \textit{rational} decisions; decisions
that are made rational by their justifications. Indeed, we have provided a
part of the justification by demonstrating how the structure of quantum
gambles, together with the gambling rules, dictate certain constraints on
the assignment of probability values. The trouble is that these probability
values violate classical constraints, for example Bell's inequalities. A
hundred and fifty years ago Boole had considered these and other similar
constraints as ``conditions of possible experience'', and consequently
conditions of rational choice. Today, we witness the appearance of
`impossible' experience. The Bohmian explains it away by reference to
unobservable non-local measurement disturbances. The instrumentalist, in
turn, insists that there is nothing to explain. But the violations of the
classical constraints (unlike the measurement disturbances) are provably
real. Therefore, something should be said about it if we insist that
``probability theory is simply the quantitative formulation of how to make
rational decisions''.

Instrumentalists often take their `raw material' to be the set of space-time
events: clicks in counters, traces in bubble chambers, dots on photographic
plates and so on. Quantum theory imposes on this set a definite structure.
Certain blips in space-time are identified as instances of the same event.
Some families of clicks in counters are assumed to have logical relations
with other families, etc. What we call \textit{reality }is not just the bare
set of events\textit{, it is this set together with its structure, }for all
that is left without the structure is noise. It has been von Neumann's great
achievement to identify this structure, and derive some of the consequences
that follow from its details. I believe that von Neumann's contribution to
the foundations of quantum theory is exceedingly more important than that of
Bohr. For it is one thing to say that the only role of quantum theory is to
`predict experimental outcomes', and that different measurements are
`complementary'. It is quite another thing to provide an \textit{%
understanding} of what it means for two experiments to be incompatible, and
yet for their possible outcomes to be related; to show how these relations
imply the uncertainty principle; and even, finally, to realize that the
structure of events dictates the numerical values of the probabilities
(Gleason's theorem).

Bohr's position will not suffice even for the instrumentalists. Their view,
far from being metaphysically innocent, is founded on an assumption which is
more radical than that of the hidden variable theories. Namely, the taxonomy
of the universe expressed in the structure of the set of possible events,
the quantum gambles which are made possible and the theory of probability
they imply, are new and only partially understood pieces of knowledge. It is
the task of an interpretation of quantum mechanics to make sense of these
structures and relate them to what we previously used to call `probability'
and even `logic'\footnote{%
See Demopoulos, 2002 for an attempt at such an explenation.}.

\bigskip

\bigskip

\begin{center}
\bigskip

\bigskip

\bigskip

\textbf{References}
\end{center}

\bigskip

Barnum, H. Caves, C. M. Finkelstein, J. Fuchs, C. A. and Schack, R. (2000)
Quantum Probability from Decision Theory? \textit{Proceedings of the Royal
Society of London A 456}, 1175-1182.

Bell, J. S. (1964) On the Einstein -Podolsky-Rosen Paradox. \textit{Physics
1, }195-200.

Beller, M. (1999) \textit{Quantum Dialogue.} Chicago: The University of
Chicago Press.

Birkhoff, G. and von Neumann, J. (1936) The Logic of Quantum Mechanics. 
\textit{Annals of Mathematics 37,} 823-843.

Bub, J. (1974) \textit{The Interpretation of Quantum Mechanics.} Dordrecht:
Reidel

Bub, J. (1997) \textit{Interpreting the Quantum World.} Cambridge: Cambridge
University Press.

Bub, J. and Clifton, R. (1996) A Uniqueness Theorem for ``No Collapse''
Interpretations of Quantum Mechanics. \textit{Studies in the History and
Philosophy of Modern Physics 27}, 181-219.

Bub, J. Clifton , R. and Goldstein S. (2000) Revised Proof for the
Uniqueness Theorem for ``No Collapse'' Interpretations of Quantum Mechanics. 
\textit{Studies in the History and Philosophy of Modern Physics 31}, 95-98.

Caves, C. M. , Fuchs, C. A. and Schack, R. (2002) Quantum Probabilities as
Bayesian Probabilities. \textit{Physical review A 65}, 2305, 1-6.

de Finetti, B. (1972) \textit{Probability Induction and Statistics.} London:
John Wiley and Sons.

Demopoulos, W. (1976) The Possibility Structure of Physical Systems. In W.
Harper and C. A. Hooker (eds.) \textit{Foundations and Philosophy of
Statistical Theories in the Physical Sciences.} Dordrecht: Reidel.

Demopoulos, W. (2002) Two Notions of Logical Structure and the
Interpretation of Quantum Mechanics. (Unpublished manuscript).

Einstein, A. Rosen, N. and Podolsky, B. (1935) Can Quantum-Mechanical
Description of Physical Reality be Considered Complete? \textit{Physical
Review 47}, 777-780.

Finkelstein, D. (1962) The Logic of Quantum Physics. \textit{Transactions of
the New York Academy of Sciences 25}, 621-637.

Fuchs, C. A. (2001) Quantum Mechanics as Quantum Information (and Only a
Little More) \textit{quant-ph/0205039}

Fuchs, C. A. and Peres, A. (2000) Quantum Theory Needs No Interpretation 
\textit{Physics Today, }March.

Greenberger, D. M. Horne M. A. and Zeilinger A. (1989) Going Beyond Bell's
Theorem. In M. Kafatos (ed) \textit{Bell's Theorem Quantum Theory and
Conceptions of the Universe.} Dordrecht: Kluwer.

Gleason, A. M. (1957) Measures on the Closed Subspaces of a Hilbert Space. 
\textit{Journal of Mathematics and Mechanics 6, }885-893.

Hardy, L. (2001) Quantum theory from Five Reasonable Axioms. \textit{%
quant-ph/0101012.}

Kernaghan, M. (1994) Bell-Kochen-Specker Theorem with 20 Vectors. \textit{%
Journal of Physics A 27} L829.

Kochen, S. and Specker, E. P. (1967) The Problem of Hidden Variables in
Quantum Mechanics. \textit{Journal of Mathematics and Mechanics 17}, 59-87.

Mermin, N. D. (1990) Simple Unified Form for the Major Unified
No-Hidden-Variables Theorems. \textit{Physical Review Letters }65, 3373-3376.

Pagonis, C. and Clifton, R. (1995) Unremarkable Contextualism: Dispositions
in Bohm's Theory. \textit{Foundations of Physics 25}, 281-296.

Pitowsky, I. (1989) \textit{Quantum Probability Quantum Logic}. Lecture
Notes in Physics 321. Berlin: Springer.

Pitowsky, I. (1994) George Boole's ``Conditions of Possible Experience'' and
the Quantum Puzzle. \textit{British Journal for the Philosophy of Science 45}%
, 95-125.

Pitowsky, I. (1998) Infinite and Finite Gleason's Theorems and the Logic of
Uncertainty. \textit{Journal of Mathematical Physics 39}, 218-228

Pitowsky, I. (2002) Range Theorems for Quantum Probability and Entanglement.
In A. Khrennikov (ed) \textit{Quantum Theory: Reconsideration of
Foundations. }Tokyo: World Scientific.

Pitowsky, I. and Svozil, K. (2001) New Optimal Tests of Quantum Non-Locality 
\textit{Physical Review A 64}, 4102-4106.

Putnam, H. (1965) Philosophy of Physics. In\textit{\ Mathematics Matter and
Method - Philosophical Papers Volume1. }Cambridge: Cambridge University
Press 1975\textit{.}

Putnam. H. (1968) The Logic of Quantum Mechanics. In\textit{\ Mathematics
Matter and Method - Philosophical Papers Volume1. }Cambridge: Cambridge
University Press, 1975.

Ramsey, F.P. (1926) Truth and Probability. In D. H. Mellor (ed) \textit{F.
P. Ramsey: Philosophical Papers}. Cambridge: Cambridge University Press 1990

Savage, L.J. (1954) \textit{The Foundations of Statistics}. London: John
Wiley and Sons

Schack, R. Brun, T. A. and Caves, C. M. (2001) Quantum Bayes Rule. \textit{%
Physical Review A 64}, 014305 1-4.

Svozil, K.(1998) \textit{Quantum Logic} Singapore: Springer

Vallentini, A.(1996) \textit{Pilot-Wave Theory of Physics and Cosmology}.
Cambridge: Cambridge University Press.

von Neumann, J. \textit{Mathematical Foundations of Quantum Mechanics.}
Princeton: Princeton University Press. (German edition 1932).

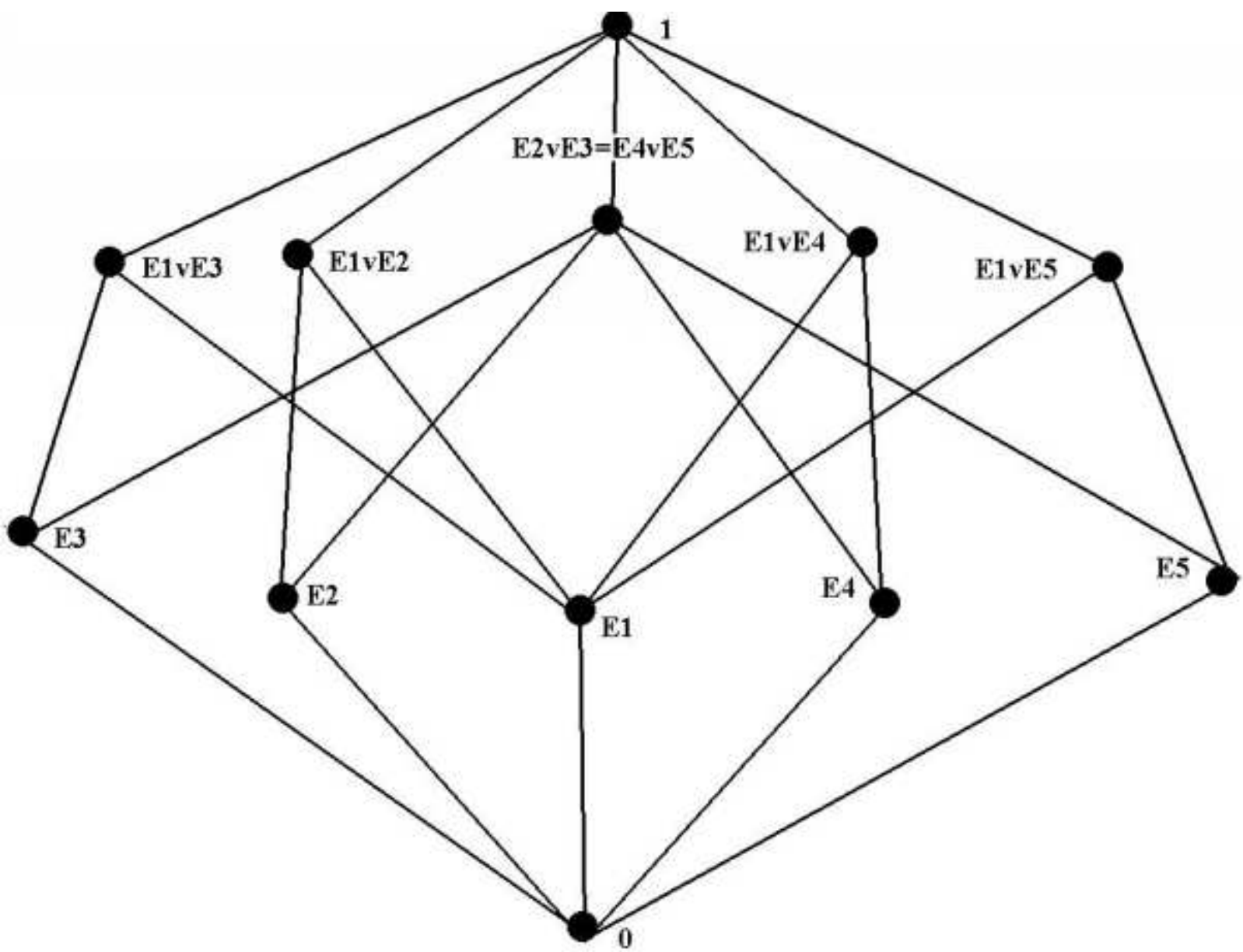
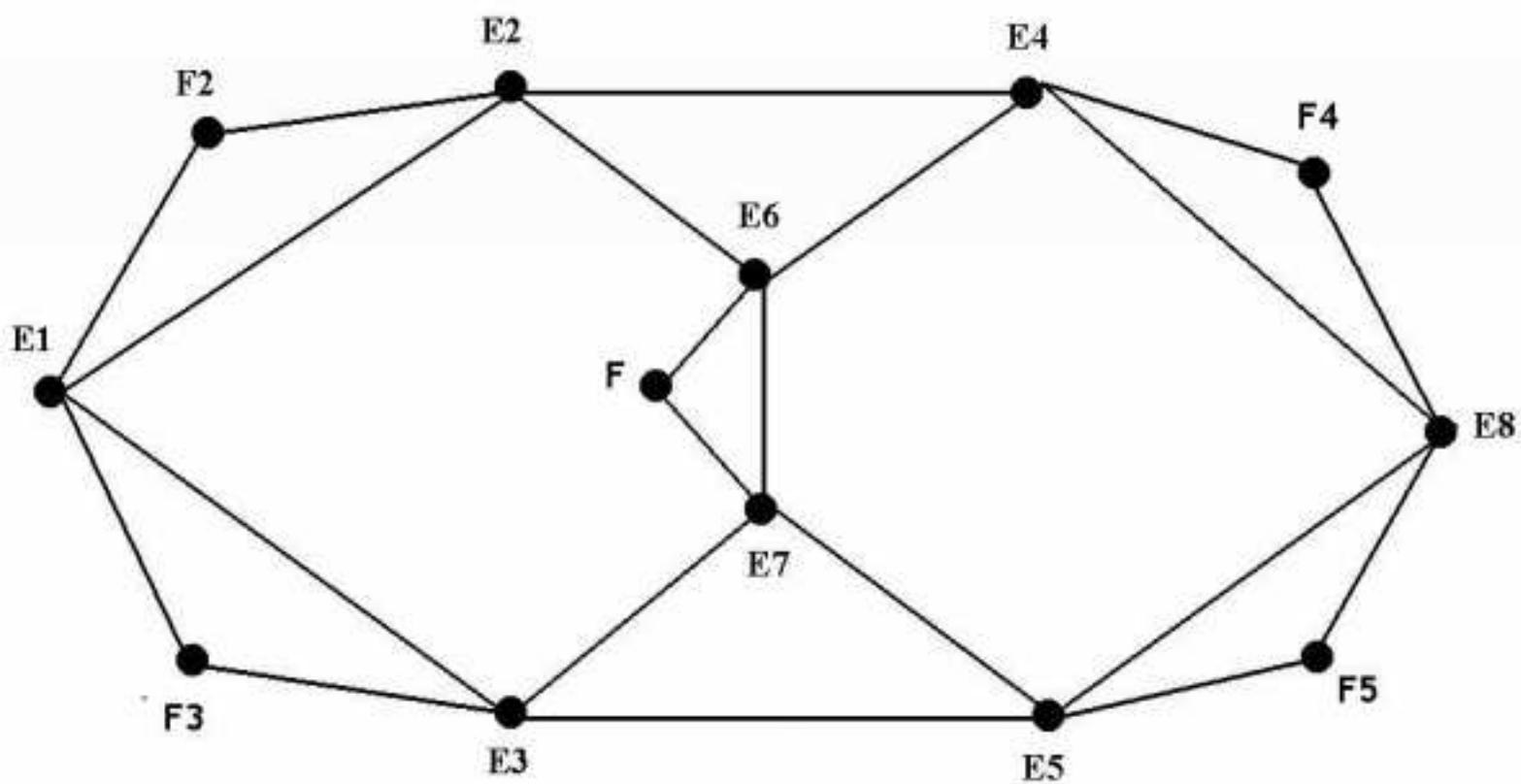
\end{document}